# The AMPEL Platform: A Concept for Open-Source Collaboration in Developing In-House Medical Device Software under EU-Regulation

**Daniel Steinbach[1†*], Maria Schmidt[1†], Alexander Twrdik[1], Clemens Malkawi[1], Toralf Kirsten[2], Martin Federbusch[1]**

1 University of Leipzig Medical Center, Department for Clinical AI and Translational Medicine, Leipzig, Germany
2 University of Leipzig, Department Medical Data Science, Leipzig, Germany
† These authors contributed equally to this work.
* ampel@medizin.uni-leipzig.de

## Abstract

Clinical decision support systems (CDSS) are expected to enhance medical treatment through automated data analysis, yet a substantial evidence gap persists as to whether they actually improve patient outcomes. It arises due to the black-box nature of commercial products and the constrains of the European medical device regulation (MDR) for such studies: patient outcomes may only be collected through the use of medical device software (MDSW). In-house MDSW under Article 5(5) MDR is a promising solution, allowing healthcare facilities to manufacture and study such software for their own use under reduced requirements. However, the overall limited experience with in-house MDSW development in Europe currently restricts the widespread use of this regulatory approach. While open-source software generally fosters evidence generation, publishing an in-house MDSW as open source would risk a classification as "placement on the market", which the MDR prohibits.

The concept presented here builds on experience from developing an in-house CDSS at the University of Leipzig Medical Center, expert discussions, a literature review and a formal legal opinion. A companion online wiki summarizes and regularly updates the results. The current concept outlines a shared open-source codebase, which doesn't constitute a medical device under the MDR. Also, "Data Input" modules must be programmed separately at each site, while still allowing for multicenter collaboration on the codebase. Atypically for open-source projects, this leaves a considerable effort to deploy the software within each site, e.g. each healthcare facility must independently meet the MDR requirements for an in-house MDSW.

The AMPEL concept describes a regulatory pathway for the open-source publication and development of in-house MDSW. It underpins the open-source AMPEL platform, which helps healthcare facilities build their own in-house CDSS.

## Plain language summary

Software can support medical decision-making. In order to minimise potential risks in the context of patient care, it must comply with specific regulatory requirements. An unintended side effect is that studies of its impact on patient outcomes are so demanding that they are rarely conducted. This work presents a concept that builds on collaboration within an open-source project and enables healthcare facilities to meet these requirements collectively and more quickly, thereby establishing the foundation for such studies to be carried out more frequently.

## Introduction

Patient care requires daily decision-making across diverse settings, ranging from primary care to emergency and inpatient care. Key challenges include severe time constraints and fragmented or incomplete information. With advances in the availability of data and models, great hopes are being placed on "AI support" for diagnosis and treatment in patient care. The number of market-available Clinical Decision Support Systems (CDSS) has increased significantly in recent years. At the same time, there are hardly any studies on their effectiveness. Fewer than 1% of these medical devices have evidence on patient outcomes.[1,2] At the same time, two of the six most frequently reported influencing factors for the implementation and adoption of CDSS are “usefulness/benefits” and “trust”.[3] Also, the closed-source nature of the products prevents independent review or further development.

In Europe two research strands can be identified: theoretical research on clinical decision support with few regulatory hurdles and real-world investigations within patient care, which are bound by the requirements for medical device software (MDSW; the MDR equivalent of 'software as a medical device', SaMD) under the medical device regulation (MDR). Studies of the second kind are expensive, time-consuming, and have uncertain outcomes in terms of marketing usability; moreover, they are not required for MDR approval in the risk class applicable to CDSS.[4] This leads to little to no incentive for manufacturers to conduct these clinical studies.

While the reproducibility of an open-source MDSW alternative would clearly enable independent review, especially addressing bias control[5,6], it remains unclear whether it would generate stronger clinical evidence. This lack of information is primarily due to limited availability of open-source MDSW in the world. Tidepool is the only example of such software to date. However, the experience with it is promising: together with its open-source community products, it has been the subject of an exceptionally high number of clinical studies.[7–11] In addition, research shows that open source can generally lead to greater trust, collaboration, and reproducibility.[12–14]

A manufacturer-independent alternative for conducting clinical studies is to use in-house MDSW under Article 5(5) of the MDR, which offers a pathway with reduced regulatory requirements. Combined with the resource efficiency of open-source code reuse, particularly in already CDSS-specific projects, time and costs could potentially be saved. Furthermore, an open-source release could facilitate multi-center development (improvements and extensions)[14] - all leading to a promising collaborative solution for generating the missing evidence. However, an in-house MDSW cannot simply be released as open source, as this would currently be interpreted as an impermissible “placing on the market” under the MDR (Art. 5, MDR) and, to the best of our knowledge, no in-house MDSW has yet been released as open source in Europe.

Besides the need for practical, reusable source code that healthcare facilities can draw on, there are few resources for developing an understanding of in-house MDSW.[15] If a healthcare facility wants to develop its own CDSS, there is, to our knowledge, no public example showing source code or completed technical documentation. Consequently, openly accessible training resources are needed to qualify more experts who can do both, understand the applicable regulatory requirements and derive practical solutions for their in-house MDSW.

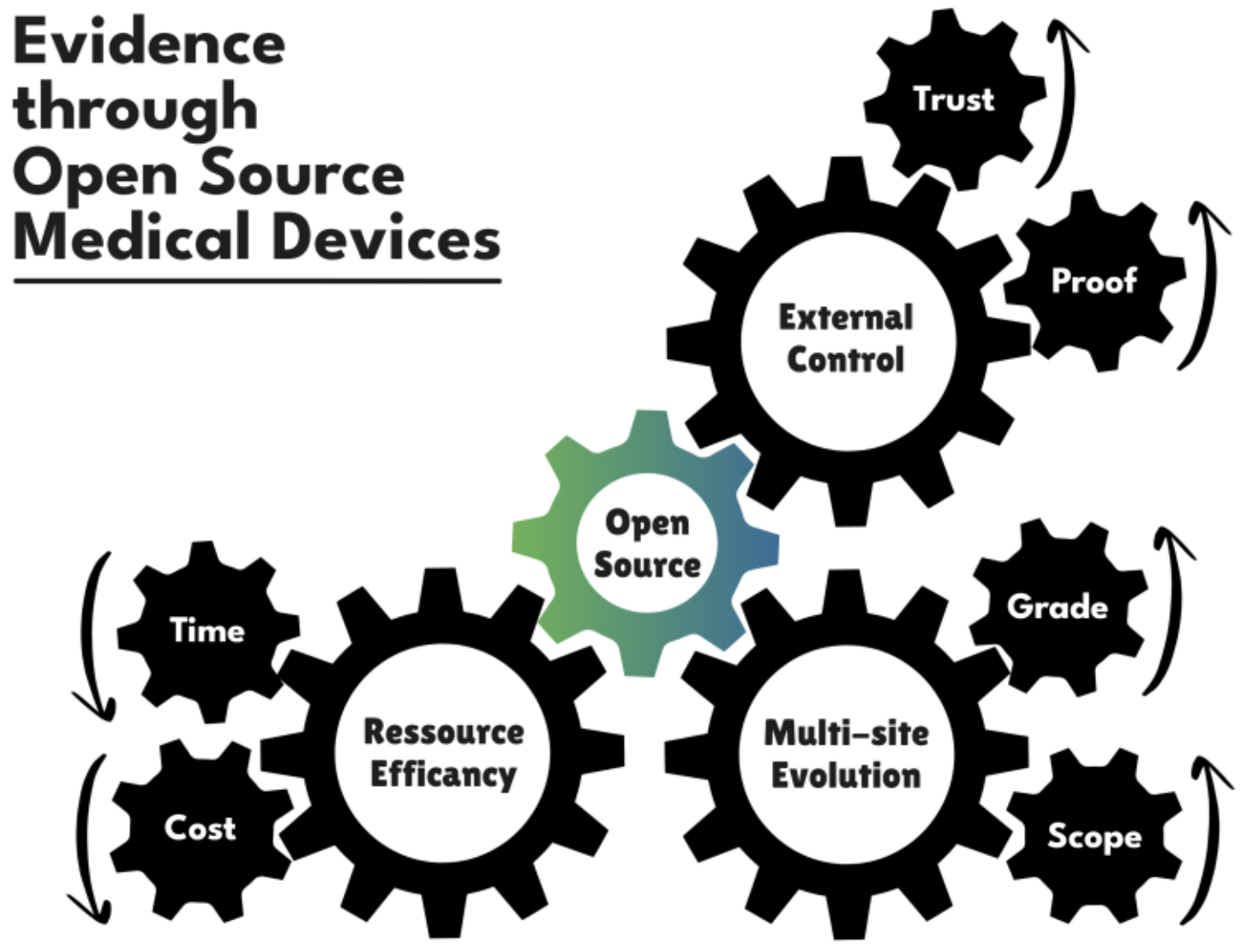

**Figure 1.** Potential impact of open source medical devices on evidence generation. Leveraging public preliminary work reduces the resources required for software development, saving both time and money. Consequently, research projects become realistically achievable within funding periods or can be completed faster. Multi-site development accelerates software evolution in both quality and quantity. Full reproducibility allows external control to take place without constraints, thereby building additional evidence and trust.

## Motivation and objective

| | |
|---|---|
| AMPEL Platform | Open-source software, its technical documentation and instruction manuals. |
| AMPEL Concept | The platform's underlying concept for medical device development, publication, and multicentric collaboration. |
| AMPEL Wiki | Web-based publication of all text documents associated with the platform and the concept. |

The AMPEL platform is intended to answer the challenges outlined above and provides hospitals and research facilities with an open-source software, together with its technical documentation and instruction manuals, for developing an in-house CDSS. The AMPEL concept presented in this paper aims to show the development of the corresponding regulatory pathway and how it can be applied by others. In particular, the concept seeks to demonstrate how the strengths of both approaches — in-house MDSW and open source — can be combined.

When LAMPE, an in-house medical device implementation of the AMPEL platform, was put into service in early 2026, the University of Leipzig Medical Center (ULMC) became the first hospital in Europe to use such an in-house CDSS.[16] Our motivation is to enable other sites to implement an in-house CDSS in a comparable way and to support them in conducting their own interventional studies that generate evidence on patient outcomes.

## Concept Development Process

The concept described here is the result of a four-year effort to transition the original CDS research project AMPEL (2019–2022) into a sustainable multicenter platform. Accordingly, the following sections describe the key steps in developing the concept.

### Governance

Following the project's successful completion, plans for a conventional spin-off were set aside in favor of establishing the project as a new department within the ULMC.  Although the goal of deploying a CDSS in routine care was similar to that of commercialization, from the outset the focus was more on maximizing scientific evidence rather than profits. This course of action called for a new regulatory concept and the creation of a research platform that provides centralized access for future participating sites.

Discussions were held in various formats with internal and external experts in medical informatics, medical law, clinical decision support, biomedical engineering, quality and risk management, insurance, data protection, hospital IT, ethics, and startup initiatives. We also exchanged information with the community of Tidepool - to date the only open-source medical device software with FDA clearance.[9] Also, the Clinical Decision Support Network (CDSN) was founded to foster interdisciplinary exchange. Frequently asked questions from the community were collected and used to continuously update an internal wiki.

### Development of the in-house medical device software (MDSW)

The concept was started and then further developed on lessons learned during development and operation of the first in-house MDSW based on the AMPEL platform, called LAMPE. Only this real-world experience could reveal which parts of the concept were feasible and which were not.

#### *Extension of the healthcare facility's Quality Management System (QMS)*

The development and operation of MDSW requires documentation within a QMS, where QMS templates and standard operating procedures (SOPs) are followed. Specific elements needed to be added to ULMC's existing ISO 9001-certified QMS for the development of in-house MDSW.[17] The Institute for Computer-Assisted Surgery (ICCAS) at Leipzig University's Faculty of Medicine was commissioned to provide consulting services for this step. An ISO 13485 certification of the QMS itself was not required.

#### *Product development within the QMS*

Experience was drawn from the research project's previous software (2019–2022), but the AMPEL platform was developed from ground up. In addition to numerous technical innovations, this made it possible to develop the software entirely in accordance with the regulatory requirements for in-house MDSW. In practical terms, the QMS templates were completed and the SOPs followed. All device-specific QMS documents were created by the newly established ULMC Department for Clinical AI and Translational Medicine.

## Laying the groundwork for multicenter collaboration

*Literature and concept draft*

In light of ongoing changes to interpretations of the MDR and the AI Act, as well as announced revisions to both, we continuously consulted relevant literature sources for updates using the following search terms and topics: (in-house) MDSW under the MDR and AI Act, “clinical investigations” and "placement on the market" under the MDR, open-source medical devices, data protection impact assessments under the GDPR, and the definition of AI under the AI Act. Particular attention was given to the legislative texts themselves and the guidance documents directly associated with them (e.g. MDCG guidance), in order to develop the regulatory pathway for a possible open-source release of the AMPEL platform.

*Formal legal opinion*

The final concept for the open-source release was prepared and reviewed through a paid legal opinion. In particular, the review addressed the “placing on the market of in-house MDSW” through open-source publication, potential licenses and disclaimers, and the use of the AMPEL platform from the perspective of an external healthcare facility.

*Development of the AMPEL Wiki*

With an emphasis on version control and user-friendliness, the technical implementation of the necessary online wiki was developed. Technical and regulatory how-to manuals were created. The QMS documents of LAMPE, originally prepared in German in a proprietary format, were translated into English and formatted in markdown for easy version control, and site-specific content was replaced with generic placeholders. Following an internal review, the documents are published in the same git repository as the technical source code. The final wiki was divided into three chapters, each aimed at a different audience: a technical manual for software developers, a regulatory manual for project managers, and an evidence repository for researchers and clinical users.

## Ethics statement

Ethical approval was not required for this study, as it did not involve human participants, animal subjects, or identifiable personal data.

## Consent statement

Consent is not applicable, as this manuscript does not contain any individual data in any form.

## Study registration

Study registration is not applicable.

## The resulting AMPEL concept for multicentric collaboration

### The AMPEL platform

If you picture an MDSW as an assembled LEGO set, the AMPEL platform can be thought of as building blocks accompanied by an instruction manual, with one major block intentionally left out. Except for this limitation, the platform provides all essential components required to develop a hospital-wide Clinical Decision Support System (CDSS) for patient care (Figure 2). Dividing the overall system into Data Input, Calculation, and Data Output, the open-source publication excludes the first component. Consequently, deployment sites are responsible for independent development of their Data Input module. Alongside technical implementation, regulatory compliance as an in-house medical device must be established. The AMPEL platform goes further than supplying source code and supports this integration by providing a product-specific regulatory manual and completed QMS documents that can serve as guiding examples or templates.

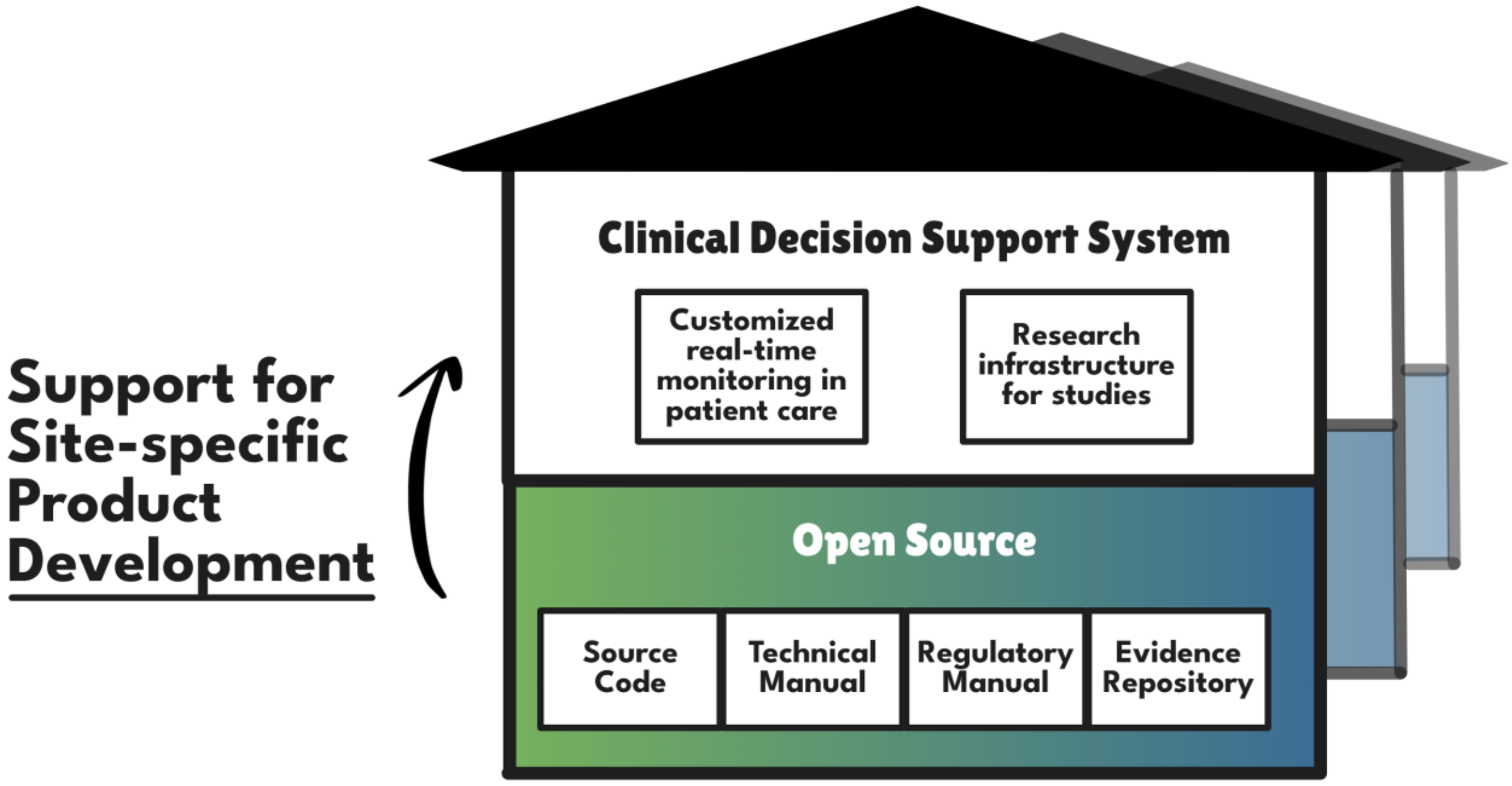

**Figure 2.** The AMPEL platform as the basis for developing clinical decision support systems as in-house medical devices.

### Benefits of in-house medical devices

The concept is based on two key assumptions. First, users intend to customize the platform for their sites. Second, they plan to conduct clinical decision support (CDS) studies. Compared with the alternative of using a CE-marked medical device (similar to Tidepool), the in-house approach enables faster development cycles and allows studies to be conducted sooner:

- Reduced requirements for the quantity and depth of QMS documentation.
- No approval is needed from a notified body (e.g. TÜV).
- No approval is needed from the national competent authority (e.g. BfArM) for interventional studies within the intended purpose of the device.[18]
- No approval is needed from an external manufacturer for interventional studies outside of its intended purpose or with additional invasive or burdensome procedures.

In principle, in-house production gives a healthcare institution maximum independence. It relies neither on a company nor on a notified body or any other regulatory authority. In-house production thus creates precisely the conditions a site needs to generate evidence independently. Figure 3 compares the requirements and benefits of in-house and CE-marked MDSW.

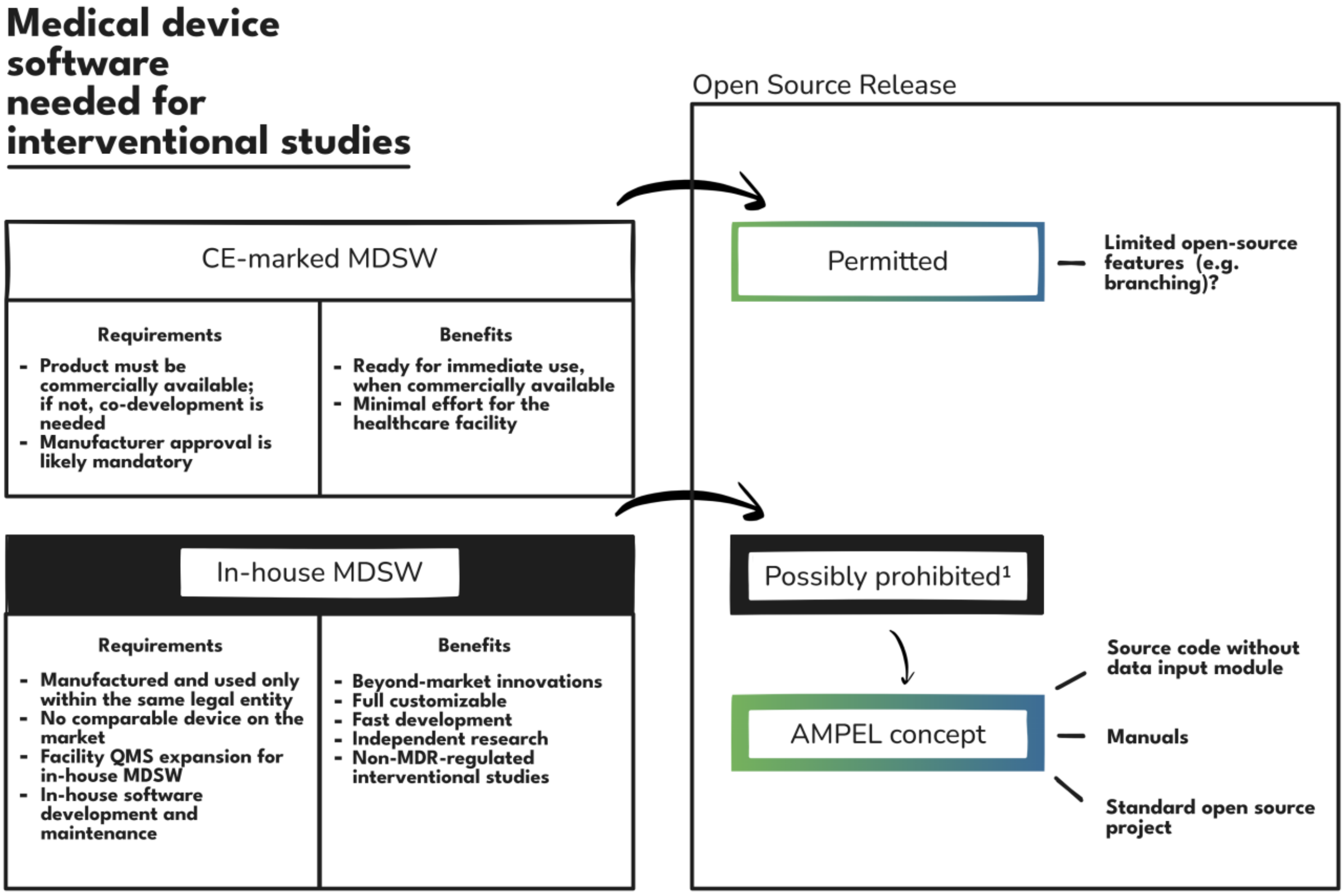

**Figure 3.** Comparison of in-house and CE-marked medical device software (MDSW) regarding their benefits for interventional studies and open-source release. Black borders indicate the path of the AMPEL platform. [1]Simplified - whether an open-source publication always constitutes a “placement on the market” is a matter of ongoing debate.

## Open-source release of in-house medical device software

Placing an in-house medical device on the market revokes its in-house status, and the healthcare institution would be required to cease operating the software. Current regulation suggests that open-source releases of MDSW could be interpreted as such a placement on the market. The AMPEL regulatory concept safeguards against this through three key mechanisms:

1. The published AMPEL platform is not a medical device. It does not include a data input module. As a result, a critical component required to run the software is absent. The remaining code represents a partial product and, without a medical purpose, is not a medical device under the MDR.
2. Without the capability to process medical input data, the platform objectively falls outside the definition of "software" under the MDR.
3. Furthermore, the release is not carried out "in the course of a commercial activity" as specified in the MDR's definition of "placing on the market".

How is the scope of the medical intended purpose (1) defined? This distinction is most effectively illustrated through the targeted user groups. The AMPEL platform is intended for researchers and software developers. They can develop a product targeting medical professionals with a medical intended purpose. Likewise, researchers and software developers may choose to utilize the software for non-medical intended purposes, such as recruiting patients for clinical trials.

Doesn't all source code have some kind of data input (2)? The MDCG has defined "data input" as follows: "Data received from/transmitted by devices.".[19] No device can provide data in the unique AMPEL format. Consequently, without a data input module, there can be no "processing of input data," and the MDR definition of "software" does not apply to the published source code.

Surely "commercial activity" (3) is a broad term – can't non-profit organizations carry out such activities too? Indeed, there are no clear rules, and the matter must be decided on a case-by-case basis. The Blue Guide specifies the criteria to be used in making such a determination: "This can only be appreciated on a case by case basis taking into account the regularity of the supplies, the characteristics of the product, the intentions of the supplier".[20] Although the source code can be downloaded at any time, its research focus, the product's incompleteness, and the use of the permissive MIT license rule out treating it as a "commercial activity."

While technically a single key mechanism would suffice, the AMPEL concept differs from "placement on the market" in a number of ways and provides a safe regulatory basis for the open-source release. More details can be found in the formal legal opinion (original in German) included in the online wiki.

## Tasks for sites adopting the platform

### *General requirements*

If another site wishes to use the AMPEL platform to develop a CDSS as an in-house medical device, it must first meet the general requirements. In brief, Article 5(5) of the MDR specifies that in-house development is permitted only if the following conditions are met:

1. The device is manufactured and used only within the same legal entity.
2. There is no comparable device in purpose or performance on the market.
3. The health facility uses an appropriate quality management system

It must therefore be ensured that the development department is in fact part of the healthcare facility's legal entity. Particularly at sites where the university and university hospital are separate entities, this may require existing experts to transfer from the university to the university hospital. Second, as part of a market analysis, it must be established that no product is already available on the market that serves the same intended purpose and meets the required performance. The third general requirement is that the healthcare facility must have defined processes and templates in its quality management system that are sufficient to satisfy the requirements for in-house medical devices. In most cases, this means extending an existing QMS (e.g., one based on ISO 9001) with additional documents. ISO 13485 certification, although often cited, is not required and should be carefully considered in light of the additional investment and expenses involved.

| Required task for in-house development | Level of supporting content | Platform documents |
|---|---|---|
| ***General requirements*** | | |
| Project management | Low | - Regulatory manual |
| Market analysis | high | - QMS document “Market analysis”<br>- Evidence paper<br>- Regulatory manual |
| Facility QMS expansion for in-house medical devices | low | - Exemplary QMS documentation<br>- Regulatory manual |
| ***Software development*** | | |
| Development of data input part | none | |
| Integration as SOUP or customized development | high | - Technical paper<br>- Technical manual<br>- Exemplary QMS documentation<br>- Validation data sets<br>- Open-source community |
| ***Technical Documentation*** | | |
| Complete, site-specific QMS documentation | medium | - Regulatory manual<br>- Exemplary QMS documentation |
| Data protection & IT security documents | low | - Regulatory manual |
| Check for AI Act requirements | medium | - Regulatory manual |
| ***Research*** | | |
| Study design | medium | - Regulatory manual |
| Cumulative Research | low | - Evidence repository<br>- Dissemination support |

**Table 1.** Required tasks and level of support for the construction of an in-house medical device based on the AMPEL platform at the time of publication.

*Product-specific requirements*

If the general requirements are met, development of a CDSS based on the AMPEL platform can begin. The AMPEL platform provides varying levels of support for the necessary tasks (see table 1). We expect that software integration in particular requires substantial support, and have therefore prioritized it here. In principle, integration can follow one of two paths:

1. Integration as "software of unknown provenance" (SOUP), which requires less development and documentation effort but offers limited customization.
2. Parts of the source code are copied and used as in-house code (Copy&Paste). Documentation and testing must be carried out independently.

The AMPEL Wiki, published alongside this paper, includes a technical and regulatory instruction manual and the beginnings of an evidence repository: https://ampel-wiki.eu

## Discussion

### Comparable projects

To date, only a few healthcare facilities have put in-house MDSW into service. One possible reason is that requirements for the form and content of QMS documents are perceived as too demanding. For example, ISO 13485 certification is frequently but erroneously described as a prerequisite for in-house MDSW.[21–23] While this certification meets the requirements for CE-marked and in-house devices alike, it sets standards for in-house developments that go beyond those specified in the MDR – which makes it unnecessarily burdensome.

Pursuing the highest possible standards may initially seem ethical. Yet, the widespread practice of making no concessions compared with CE-marked devices often means that in-house MDSW are not developed at all[24,25]. In the authors' opinion, it also leads to the other extreme; the claim that a software is in fact not an MDSW and requirements can be fully ignored. We see a promising approach to resolving this issue in publishing real-world technical documentation for in-house MDSW, explicitly highlighting critical measures while showing acceptable simplifications relative to CE-marked devices.

This study identified two public projects that support healthcare facilities preparing technical documentation for MDSW. In 2025, the Clinical Scientific Computing Team at Guy's and St Thomas' NHS Foundation Trust in London published open-source QMS templates.[26] An example repository containing completed documentation has been announced. Although the service is not publicly available, the team also advises other NHS trusts on how to develop MDSW. It should be noted that, following Brexit, the UK did not implement the latest revisions of the MDR and has instead taken a separate national approach.

While this NHS project can primarily help establish a foundational QMS, the open-source project Tidepool provides real-world technical documentation together with the source code.[7,27] The software for automated insulin pumps, which received FDA clearance in 2023, is designed for direct patient use. In addition to its different target audience, it does not address the European regulatory framework (MDR, AI Act, GDPR) or the specific circumstances of in-house MDSW.

### Open-source publication

The open-source release of Tidepool would be equivalent in Europe to the release of a CE-marked device. A major advantage is that it could be used without additional steps. Unlike in-house devices, placing the product on the market is explicitly permitted in this case. However, without a European example, several questions remain. To what extent can the mechanisms of an open-source project be used? What happens, for example, if an independent developer creates a publicly accessible branch? What must healthcare institutions and companies consider if they want to use CE-marked open-source medical device software? The commercially available MDSW twiist, which is based on Tidepool and was authorized by the FDA in 2024, shows that this is no longer a theoretical question. This points to a promising area for future research.

One advantage of the AMPEL regulatory concept is that it circumvents these unresolved questions in Europe by creating the conditions for a “regular” open-source project. Although the QMS documentation is updated centrally, somewhat slowing development compared with other open-source projects, the community retains full access to all open-source practices. This makes it possible to leverage the scaling and reproducibility effects that have driven innovation in other fields.

In addition, compared with a CE-marked device, the resulting in-house medical device offers substantially greater flexibility for site-specific development, as well as the possibility of conducting interventional studies without the MDR requirements for a “clinical investigation”[18], which often prevent studies from being undertaken. All this, however, entails considerable work for healthcare institutions. In addition to their obligation to be innovative relative to products available on the market, they must program their own data input and prepare site-specific QMS documentation. In summary, the AMPEL concept is demanding, but highly rewarding, particularly in scientific contexts.

### Limitations and Outlook

We view the regulatory manual as a starting point and expect it to continue to evolve. Only feedback from external sites will reveal where the regulatory manual needs to be expanded or improved. We expect a significant boost from an ongoing research project investigating specific questions about implementing the AMPEL platform at Dresden University Hospital.[28] A further limitation is that many detailed questions concerning development of in-house MDSW still need to be clarified - both at the regulatory level (e.g. through the MDCG) and in practice. In addition, the MDR is currently under revision, and changes affecting in-house medical devices can be expected. However, the proposed changes point toward a reduction in regulatory burden in this area.[29] The same applies to other components of the AMPEL platform, such as the source code, technical documentation and the evidence repository, although they involve different stakeholders. Only an active community will make the potential benefits of multicenter scaling a reality. Every feature, every algorithm, and especially every additional piece of evidence will advance CDS research and, through open-source publication, all added elements will be immediately and publicly available. We kindly invite you to join the AMPEL platform community and get involved.

## Conclusions

In-house medical device software and open-source publication are suitable approaches to making it easier for healthcare facilities to conduct research on Clinical Decision Support Systems (CDSS). By not releasing the data-input module, an in-house CDSS can be published as an open-source project, as it no longer qualifies as a medical device placed on the market.

## Data and software availability

### Software availability

The AMPEL wiki including the QMS documents, the regulatory manual and the formal legal opinion are published under the MIT license: https://ampel-wiki.eu

## AI-use declaration

A generative AI tool (“Translate”, Kagi Inc.) was used to create a first draft English translation. The translation was human-checked and improved before publication. An AI search tool (“Assistant”, Kagi Inc.) was also used as part of the literature review. Programming the online wiki was assisted by an AI tool (“Codex” and “ChatGPT”, OpenAI).

No other tasks were carried out with AI tools.

## Competing interests

No competing interests

## Acknowledgements

We thank Isabell Weber, Christian Sünkel, Johanna Martin, Katharina Käbisch, Martin Reich, Johann Berger and Thomas Neumuth for their support and collaboration in establishing the quality management documentation for LAMPE.

We thank the Executive Board of University of Leipzig Medical Center, Robert Jacob and Christoph Josten, for ensuring the long-term continuation of the AMPEL platform.

We thank Ulrich M. Gassner for the collaboration and permission to publish his formal legal opinion.

We thank all members of the clinical decision support network for their valuable input and fruitful discussions.

## Funding information

This project was co-financed by the European Union (NextGernerationEU) and through public funds according to the budget decided by the Saxon State Parliament.